# The electric field as a novel switch for uptake/release of hydrogen storage in nitrogen doped graphene

Z. M. Ao,[1,*] A. D. Hernández-Nieves,[2,3] F. M. Peeters[3] and S. Li[1]

[1] School of Materials Science and Engineering, The University of New South Wales, Sydney, NSW 2052, Australia

[2] Centro Atomico Bariloche, 8400 San Carlos de Bariloche, Rio Negro, Argentina

[3] Departement Fysica, Universiteit Antwerpen, Groenenborgerlaan 171, B-2020, Belgium

Nitrogen-doped graphene was recently synthesized and was reported to be a catalyst for hydrogen dissociative adsorption under a perpendicular applied electric field ($F$). In this work, the diffusion of H atoms on N-doped graphene, in the presence and absence of an applied perpendicular electric field, is studied using density functional theory. We demonstrate that the applied field can significantly facilitate the binding of hydrogen molecules on N-doped graphene through dissociative adsorption and diffusion on the surface. By removing the applied field the absorbed H atoms can be released efficiently. Our theoretical calculation indicates that N-doped graphene is a promising hydrogen storage material with reversible hydrogen adsorption/desorption where the applied electric field can act as a switch for the uptake/release processes.

[*] Corresponding author. zhimin.ao@unsw.edu.au

1. **Introduction**

Hydrogenation of carbon materials has been attracting a wide range of interests as a new mechanism of hydrogen storage and for tuning its electronic behaviour.[1] Graphene with its unique structural, electronic, thermal, and mechanical properties, has been identified as a promising candidate for hydrogen storage and for the next generation of electronics.[2] In most of the reported works, hydrogen is stored in molecular form,[3] although some works have considered storing hydrogen in atomic form, where H atoms were chemically bonded with the carbon atoms.[1,4,5] Due to the strong $sp^2$ bonding between C atoms in graphene as well as the π bonding between $p_z$ orbitals of the C atoms, the $H_2$ dissociation and chemical adsorption on graphene has a reaction barrier between 3.3 and 6.1 eV, depending on the adsorption site of the H atoms.[5] Most recently, it was found that the dissociation of hydrogen molecules could be induced by a perpendicular electric field $F$.[6,7] The dissociated H atoms can bind with the graphene layer and form covalent bonds. In other words, the application of an electric field can induce graphene hydrogenation.[8,9] Based on our previous work,[8] it is known that $H_2$ molecules can be spontaneously dissociatively adsorbed on N-doped graphene under a perpendicular applied electric field. These dissociated H atoms prefer to be chemically adsorbed on the C atoms near the doped N atom.

In order to be an efficient hydrogen storage material, N-doped graphene, which can be synthesized by different approaches,[10,11,12] is required to capture as much hydrogen as possible. Therefore, it is desirable that the adsorbed H atoms bonded with the C atoms near the doped N atom can diffuse to the other C atomic sites, and subsequently the adsorbed H atoms can be easily desorbed to form $H_2$ molecules once the electric field is removed. In this work, the diffusion of H atoms on N-doped graphene with and without the applied perpendicular electric field will be investigated by density functional theory (DFT). Hydrogen desorption and $H_2$ formation will also be studied to understand the release procedure of the stored hydrogen.

## 2. Computational Methodology

The DFT calculations were performed using the DMOL3 code.[13] The generalized gradient approximation (GGA) with revised Perdew-Burke-Ernzerhof (RPBE) functional was employed as the exchange-correlation functional.[14] A double numerical plus polarization (DNP) was used as the basis set, while the DFT semicore pseudopotentials (DSPP) core treatment was employed to include relativistic effects that replaces core electrons by a single effective potential. Spin polarization was considered in the calculations. The convergence tolerance of the energy was set to $10^{-5}$ Ha (1 Ha = 27.21 eV), and the maximum allowed force and displacement were 0.02 Ha and 0.005 Å, respectively. To investigate the diffusion pathways of the hydrogen atoms at the N-doped graphene surface, linear synchronous transition/quadratic synchronous transit (LST/QST)[15] and nudged elastic band (NEB)[16] tools in the DMOL3 code were used. These methodologies have been demonstrated as fantastic tools to search for the structure of the transition state (TS) and the minimum energy pathway. In the simulations, three-dimensional periodic boundary conditions were imposed, and all the atoms are allowed to relax. The supercell used to investigate the diffusion of H atoms on N-doped graphene is shown in Fig. 1. We minimized the interlayer interaction by allowing a vacuum width of 18 Å normal to the layer.

## 3. Results and discussion

Based on our previous work,[8] it is believed that a $H_2$ molecule can be dissociated and adsorbed on the C atoms near the doped N atom in graphene under an applied electric field. As H atoms are chemically adsorbed on the graphene surface, it is interesting to know the diffusion behaviour of the H atoms on graphene. This may provide new insights to enhance the hydrogen storage capacity significantly through the diffusion of absorbed H atoms. To simplify the investigation, we first consider the diffusion of a single H atom on N-doped graphene. For the H atom chemical adsorption, there are five different possible adsorption sites as shown in Fig. 1. After structure relaxation, it is found that H adsorption on the C atom at the site X, where the C atom coordinates to the doped N atom, has the lowest energy, i.e. this is the most favourable configuration for a single H atom adsorption and it is considered as the base structure for studying the diffusion of an individual H atom on N-doped graphene. The C atom bonded with the H atom is displaced from the C plane by about 0.29 Å. This value is similar to the shift of 0.32 Å that is caused by the dissociative adsorption of a hydrogen molecule on graphene.[8] This is associated with the hybridization of C atoms from $sp^2$ in graphene to $sp^3$ in graphane.[17,18] As shown in Fig. 1, there are four possible H atom diffusion pathways: X→SX, X→O, X→SX′ and X→O′. Using LST/QST and NEB calculation tools, the diffusion barriers of the four pathways are listed in Table I, which shows that all the barriers are around 2.5 eV. In addition, the diffusion increases the energy of the system. Therefore, the adsorption of a single H atom on N-doped graphene is considered to be much more stable as compared to the case of a single H atom adsorbed on pristine graphene, which has a diffusion barrier of ~1 eV.[19]

It is known that the high stability of H on N-doped graphene is not favourable for hydrogen storage due to the reversible hydrogen uptake and release requirement. As mentioned before, applying a perpendicular electric field to graphene in the presence of $H_2$ gas is an alternative approach to facilitate the hydrogenation of graphene.[8] Therefore, the diffusion behaviour under $F$ is of interest for hydrogen storage applications. Previously we found that hydrogenation occurs around $F = 0.01$ a.u. (1 a.u. = $5.14 \times 10^{11}$ V/m).[8] Table I lists the results of the diffusion along four different pathways under $F = 0.01$ a.u.. It is noted that the diffusion barrier $E_{bar}$ decreases significantly for all four pathways, and can even be negative for the diffusions of X→O and X→SX′. The energy difference between the configurations before and after diffusion $E_R$ decreases slightly as well. This evidences that hydrogen diffusion becomes favourable under the applied electric field in an ambient environment. Note that further increasing the electric field, it is expected that H can diffuse even more easily. However, larger electric field costs more energy and results in stringent requirements for the hydrogen storage system. Therefore, only the minimum electric field at 0.01 a.u. of hydrogen dissociative adsorption on N-doped graphene is considered in this work.

Fig. 2 shows the diffusion pathway of a single H atom migration from site X to site O in the presence of an electric field of 0.01 a.u.. It is clear that both configurations of initial and final states reconstruct spontaneously in the presence of the electric field. The reconstructed atomic structures of initial and final states are shown as State 1 and State 2 respectively in Fig. 2 where the H atom breaks the chemical bond with the N-doped graphene. For the TS state, the H atom keeps the weak interaction with the N-doped graphene. From Fig. 2 and Table I, we note that the reaction barrier, $E_{bar} = E_{TS}-E_{IS}$ is negative, where $E_{TS}$ and $E_{IS}$ are the energies of the transition state and the initial state, respectively. It implies that there is no energy barrier from IS to State 2. However, there is an energy difference (~0.58 eV) between State 2 and the final state (FS). Thus, the minimum energy required for the entire reaction from IS to FS is 0.29 eV as shown by $E_R$ in Fig. 2. As aforementioned, the structures of States 1 and 2 are the favourite configurations before and after diffusion under the applied electric field. For the diffusion from State 1 to State 2, the barrier, $E'_{bar} = E_{TS} - E_1$, is 0.32 eV and the diffusion energy, $E'_R = E_2 - E_1$, is 0.20 eV. Table 1 also shows $E'_R$ and $E'_{bar}$ for the other diffusion path with negative barrier. Notice that the H atom can diffuse on the N-doped graphene surface with an energy barrier $E'_{bar}$ around 0.3 eV in the presence of the applied electric field. Thus, a single H atom can diffuse throughout the N-doped graphene freely with a low energy barrier. Fig. 2 also reveals why the electric field facilitates individual H atom diffusion on the N-doped graphene surface. It shows that the H atom nearly desorbs from the N-doped graphene surface and has a weak interaction with the graphene layer due to the interaction of the electric field on the positive charged H atom. Table II shows the interaction between H and the N-doped graphene in the presence and absence of the electric field. It is found that the adsorption energy $E_{ads}$ of H atoms on N-doped graphene reduces significantly in the presence of the electric field due to the extended bond length between the positive charged H and the C of graphene. Thus, the H atom can diffuse on the surface with a low energy barrier in the presence of the electric field.

Recently, the diffusion barriers of transition-metal (TM) atoms on graphene were reported to be in the range of 0.2-0.8 eV.[20] However, if the TM adatoms are coupled to a vacancy, the diffusion barrier would increase and fall into the range of 2.1-3.1 eV.[20] A similar barrier enhancement was also predicted in the case of H atom diffusion at the graphene/graphane interface.[21] Therefore, specific defects, such as doping, vacancies and adatoms, will significantly affect the diffusive behaviour of H on graphene. In addition, in the process of hydrogenation, a $H_2$ molecule is first dissociated into two H atoms, which are then chemically adsorbed on N-doped graphene.[8] Therefore, it is important to study diffusion of two H atoms on N-doped graphene. Fig. 3 shows six possible configurations for two H atoms adsorbed on N-doped graphene. After relaxation of the structure, the configuration A3 is found to have the lowest energy. Subsequently, the different configurations are investigated under an applied electric field and it was found that the configuration A1 possesses the lowest energy when $F$ = 0.01 a.u.. However, the energy minimum sites of the two H atoms in both the A1 and A3 configurations are different from the results reported in our previous work.[8] This difference is due to the different concentration of N doping. As we are interested in hydrogenation of graphene in the presence of an electric field, the configuration A1 is taken as the initial structure for the investigation of diffusion. Table III lists the diffusion barrier and diffusion energy for the different pathways from A1 to the other five possible configurations at $F$ = 0.01 a.u. and the different pathways from these five configurations to A3 in the absence of the applied electric field. It shows that under the applied electric field, the diffusion to the configurations A2-A4 will be easier due to the lower diffusion barrier. Since the diffusion energy barrier is less than 0.31 eV, such a barrier should be easily overcome at room temperature. For the cases of diffusion to A5 and A6, the barriers are above 1.09 eV. This is because the diffusion from A1 to A5 and A6 implies a movement of both H atoms while the diffusion from A1 to A2-A4 involves a jump of only one of the two H atoms. Therefore, the diffusion from A1 to A2, A3 and A4 are clearly the most favourable pathways with both low $E_{bar}$ and $E_R$.

Due to the low barrier for H atom diffusion on N-doped graphene under the applied electric field, it is predicted that N-doped graphene can be fully hydrogenated under an applied electric field, and all the H atoms are adsorbed at single sides of graphene, i.e. this is the uptake process for the hydrogen storage material. After full hydrogenation, the electric field is removed and we study the possibility of releasing the absorbed hydrogen. Because A3 is the favourable configuration without $F$, we investigate the diffusion without $F$ from A1-A6 to A3. The results are shown in Table III. It is found that the diffusion from A1 and A6 to A3 are spontaneous with both exhibiting a negative diffusion barrier and diffusion energy. The diffusions from A2, A4 and A5 to A3 have barriers smaller than 1 eV and also negative diffusion energy. Similar to the cases of negative diffusion barriers in Table I, for the diffusions with negative barriers in Table III in the absence of the electric field, the initial structure is from the structure obtained in the presence of the electric field at 0.01 a.u.. After removing the electric field, the configuration also reconstructs. In other words, the $E'_{Bar}$ less than 0.06 eV is the actual diffusion barrier. If the reconstruction energy obtained is used for the diffusion, no additional external energy is needed, the diffusion can go through spontaneously. Note from Tables I and III that the diffusion is normally easier for the two H atoms on N-doped graphene in the absence of an electric field than for a single H atom. The explanation will be given in the last paragraph of the paper.

Therefore, once the electric field is removed, H atoms can diffuse spontaneously into the A3 configuration. For a hydrogen storage material, it is also required that the stored hydrogen can be released efficiently. The desorption process from A3 to configuration B where one $H_2$ molecule is located over N-doped graphene, as shown by the FS configuration in Fig. 4, is then studied and the results are also listed in Table III. Notice that hydrogen desorption lowers the energy, thus stabilising the system. However, there is an energy barrier of 2.12 eV which needs to be overcome for desorption. The detailed reaction coordinate is shown in Fig. 4, where the atomic structures of IS, TS and FS are also given. At TS, one H atom desorbs from graphene and binds with the other H atom. Due to the high energy barrier for hydrogen desorption, a high temperature is required to release the hydrogen atoms when the hydrogen concentration is below 0.5 wt%. This raises the question: what would happen if the hydrogen concentration is above this value? In this case, we have to consider the hydrogen desorption process in the presence of another H atom near the $H_2$ molecule in the same supercell as depicted in Fig. 5. We calculated all the possible configurations for 3 H atoms on N-doped graphene and found that the configuration shown in Fig. 5(a) has the lowest energy. If two H atoms desorb from graphene and form a $H_2$ molecule, the structure is shown in Fig. 5(b). Therefore, the desorption process occurs from the configuration shown in Fig. 5(a) to Fig. 5(b). The results are given in Table III where we found that the reaction energy and the reaction barrier are -1.24 and -1.23 eV, respectively. In other words, the stored hydrogen can be released automatically when the concentration is above 0.5 wt%.

To better understand this release process, Fig. 6 shows the detailed pathway. As shown in Fig. 6, the 2 H atoms prefer to desorb from the N-doped graphene surface and combine into a $H_2$ molecule as shown by state 1. This means that atomic H adsorption on N-doped graphene is unstable and hydrogen can be released in the form of the $H_2$ molecule spontaneously. Subsequently, this free $H_2$ molecule can diffuse on the N-doped graphene surface with a very low energy barrier (insert of Fig. 6). In other words, the highest concentration of hydrogen storage in N-doped graphene is ~0.5 wt% in the absence of an applied electric field. In the presence of the applied electric field, due to the low dissociated adsorption energy barrier for $H_2$ on N-doped graphene and the low diffusion energy barrier of H atoms on the N-doped graphene as discussed above, N-doped graphene is expected to be fully hydrogenated and the hydrogen weight ratio can reach up to 7.23 wt%, which is in excesses of the target of the U.S. Department of Energy (DOE), i.e. 6 wt%. By removing the electric field, H atoms can desorb from N-doped graphene and be released efficiently in the form of $H_2$ molecules. Therefore, the electric field is a novel and efficient switch for hydrogen storage in N-doped graphene.

From Figs. 4 and 6, we see that the presence of the third H atom on N-doped graphene induces a large difference in the energy barriers for hydrogen release. In other words, H atoms can be released spontaneously from N-doped graphene in the presence of the third H atom while a high temperature is needed to release the two H atoms if the third H atom is absent. Very recently, it was reported that the ease of adsorbate diffusion on graphene strongly depends on the carrier density of graphene.[22] Through a Mulliken analysis in this study, it is known that the H atoms transfer electrons to the N-doped graphene layer. Therefore, with the presence of the third H atom, the carrier density in N-doped graphene is higher, and the diffusion of H atoms on N-doped graphene is much easier. Similar explanation can be made for the phenomenon found above that two H atoms diffuse easier on N-doped graphene than that of only one H atom, due to the higher electron concentration from the H atoms. The lower diffusion barrier means weaker interaction between H and N-doped graphene, and the H atoms are much easier to be released. Therefore, hydrogen atoms in fully hydrogenated N-doped graphene are predicted to be released easily until the hydrogen weight ratio of 0.5 wt%, because of the higher carrier density in the N-doped graphene when the H concentration is higher. In this way, the hydrogen storage capacity of N-doped graphene is 6.73 wt%, while the electric field is the switch for the hydrogen uptake/release processes. In addition, the 0.5 wt% hydrogen left in N-doped graphene can be further removed by increasing temperature. In addition, we also consider the possibility to use pristine graphene as a hydrogen storage material. However, there is a large energy barrier for $H_2$ molecules dissociated adsorption on pristine graphene,[1,9] which prevents it to be useful for hydrogen storage applications.

## 4. Conclusion

In summary, we found that the hydrogen uptake process in a N-doped graphene layer can be facilitated by a perpendicular applied electric field, which induces hydrogen dissociative adsorption and diffusion on N-doped graphene with low energy barriers. By removing the electric field, the stored hydrogen can be released efficiently at ambient conditions when the hydrogen concentration is higher than 0.5 wt%. Therefore, N-doped graphene is a promising hydrogen storage material with storage capacity up to 6.73 wt%, and the electric field can act as a switch for the hydrogen uptake/release processes.

**Acknowledgments**

Financial supports of the Vice-Chancellor's Postdoctoral Research Fellowship Program (SIR50/PS19184) and the ECR grant (SIR30/PS24201) from the University of New South Wales are acknowledged. This work is also supported by the Flemish Science Foundation (FWO-Vl) and the Belgian Science Policy (IAP).

Table I. Diffusion barriers of several pathways (see Fig. 1) for a single H atom on N-doped graphene without electric field and in the presence of an electric field of 0.01 a.u.. (1 a.u. = $5.14 \times 10^{11}$ V/m) The energies of $E_R$, $E_{Bar}$, $E'_R$ and $E'_{Bar}$ are defined in Fig. 2. In the presence of the electric field the configurations of initial and final states reconstruct, and therefore $E'_{Bar}$ (not $E_{bar}$) is the actual diffusion barrier, see Fig. 2 and related text.

| Electric field (a.u.) | Pathway | $E_R$ (eV) | $E_{Bar}$ (eV) | $E'_R$ (eV) | $E'_{Bar}$ (eV) |
|---|---|---|---|---|---|
| 0 | X→SX | 0.61 | 2.21 | | |
| | X→O | 0.27 | 2.48 | | |
| | X→SX′ | 0.58 | 2.48 | | |
| | X→O′ | 0.66 | 2.61 | | |
| 0.01 | X→SX | 0.55 | 0.47 | | |
| | X→O | 0.29 | -0.094 | 0.20 | 0.32 |
| | X→SX′ | 0.56 | -0.10 | -0.012 | 0.266 |
| | X→O′ | 0.62 | 0.62 | | |

Table II. The adsorption energy $E_{ads}$ of H atoms on N-doped graphene in different configurations and in the presence and absence of an electric field.

| Configurations | $E_{ads}$ (eV) | |
|---|---|---|
| | $F = 0$ | $F = 0.01$ a.u. |
| X | -3.04 | -1.15 |
| SX | -2.43 | -0.60 |
| O | -2.77 | -0.86 |
| SX′ | -2.46 | -0.59 |
| O′ | -2.38 | -0.53 |
| A1 | -2.17 | -1.94 |
| A2 | -1.93 | -1.69 |
| A3 | -2.26 | -1.89 |
| A4 | -2.10 | -1.76 |
| A5 | -1.55 | -1.37 |
| A6 | -2.06 | -1.70 |

Table III. Diffusion barriers of several pathways (see Fig. 3) for two H atoms on N-doped graphene without electric field and in the presence of an electric field of 0.01 a.u.. The result for the process from A3_H to B_H (see Fig. 5) is also shown in this table. Note that similar to Table I, $E'_{Bar}$ is the actual diffusion barrier in the absence of the electric field due to the configuration reconstruction of initial and final states.

| Electric field (a.u.) | Pathway | $E_R$ (eV) | $E_{bar}$ (eV) | $E'_R$ (eV) | $E'_{Bar}$ (eV) |
|---|---|---|---|---|---|
| 0.01 | A1→A2 | 0.51 | 0.31 | | |
| | A1→A3 | 0.12 | 0.26 | | |
| | A1→A4 | 0.38 | 0.26 | | |
| | A1→A5 | 1.12 | 1.09 | | |
| | A1→A6 | 0.51 | 1.15 | | |
| 0 | A1→A3 | -0.18 | -1.55 | -0.004 | 0.02 |
| | A2→A3 | -0.66 | 1.00 | | |
| | A4→A3 | -0.33 | 0.34 | | |
| | A5→A3 | -1.38 | 0.88 | | |
| | A6→A3 | -0.39 | -1.76 | 0.012 | 0.046 |
| | A3→B | -1.39 | 2.12 | | |
| | A3_H→B_H | -1.24 | -1.23 | 0.0056 | 0.0058 |

**Figure Captions**

FIG. 1. (Color on line) Supercell of N-doped graphene used in the calculations. The grey and blue balls are C and N atoms, respectively. The letters refer to the sites of an adsorbed H atom.

FIG. 2. (Color on line) Detailed diffusion pathway of an individual H atom on N-doped graphene, where the structure of the initial structure (IS), the transition state (TS), the final state (FS) and the two energy minimum states 1 and 2 before and after TS are also given.

FIG. 3. (Color on line) Six possible configurations of 2 H atoms chemically adsorbed on N-doped graphene. The white balls are H atoms.

FIG. 4. (Color on line) Detailed reaction pathway from A3 to B, where the structure of IS, TS and FS are also given.

FIG. 5. (Color on line) The energy minimum atomic structure for (a) 3 H atoms on N-doped graphene, (b) 1 H atom and 1 $H_2$ molecule on N-doped graphene.

FIG. 6. (Color on line) Detailed reaction pathway from A3_H to B_H, where the structures of IS, TS, FS and the energy minimum state before the TS state 1 are also given.

Fig. 1

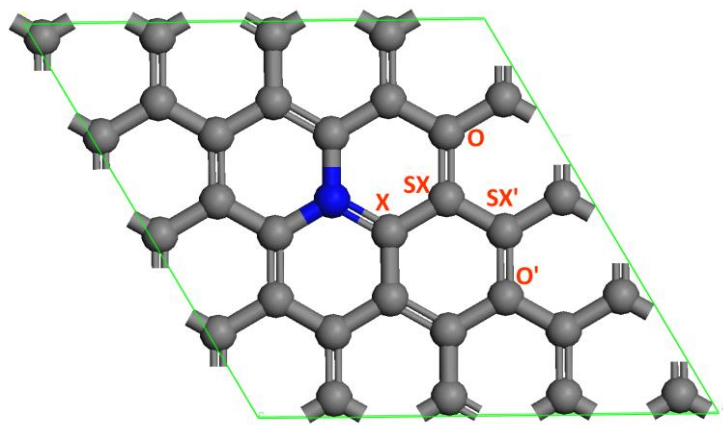

Fig. 2

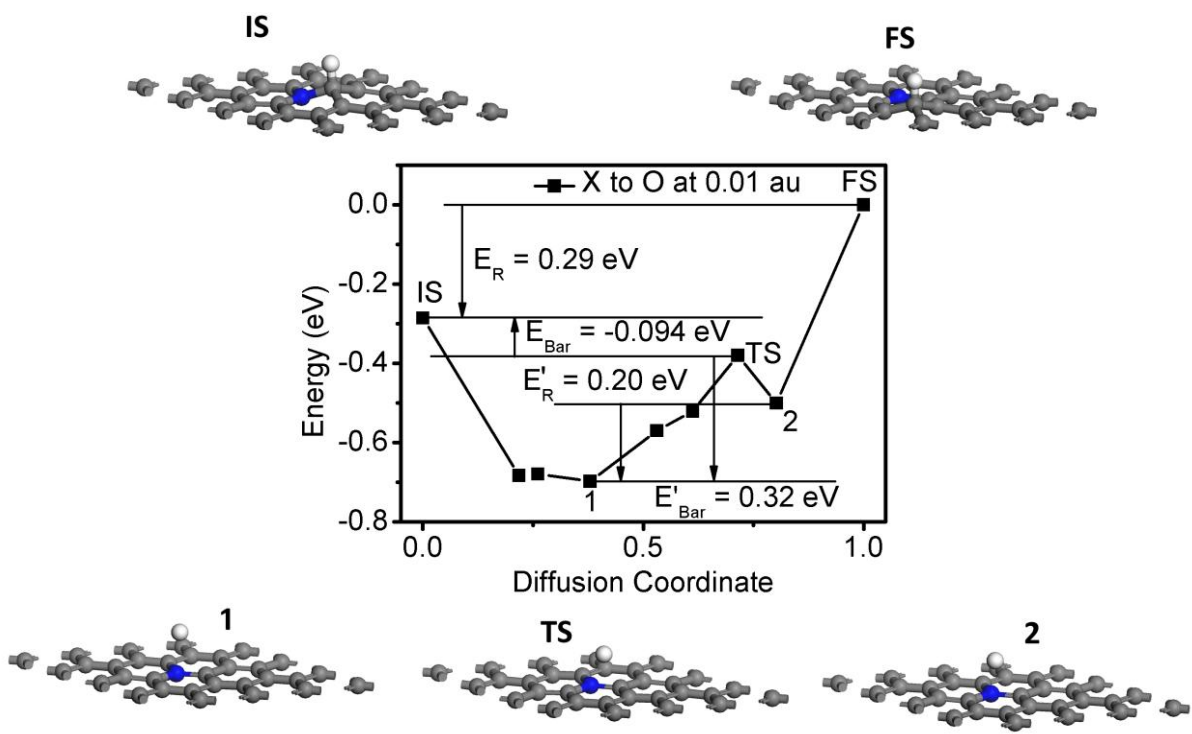

Fig. 3

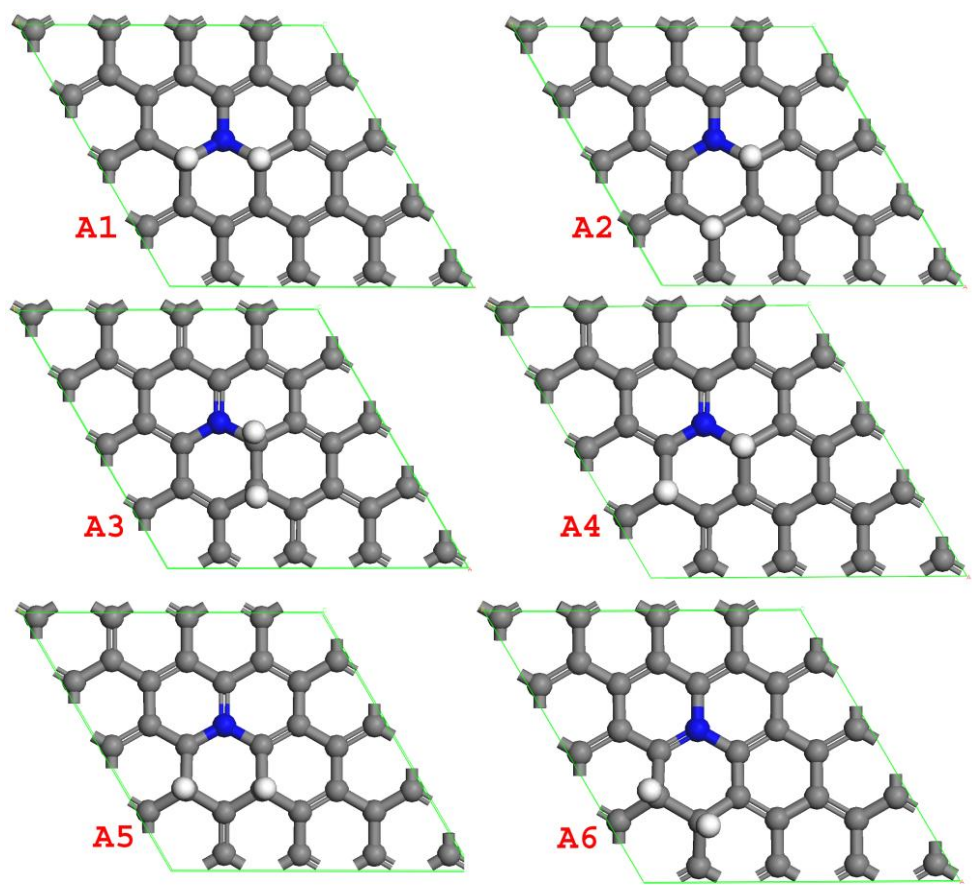

Fig.4

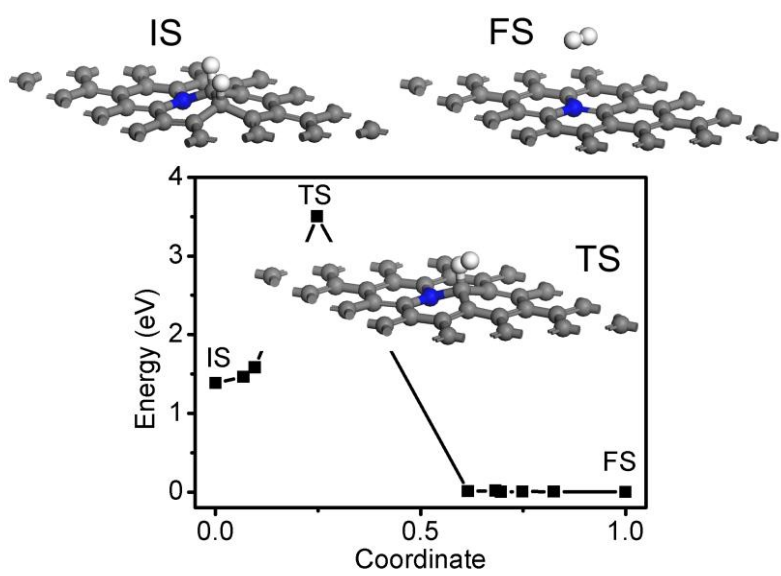

Fig. 5

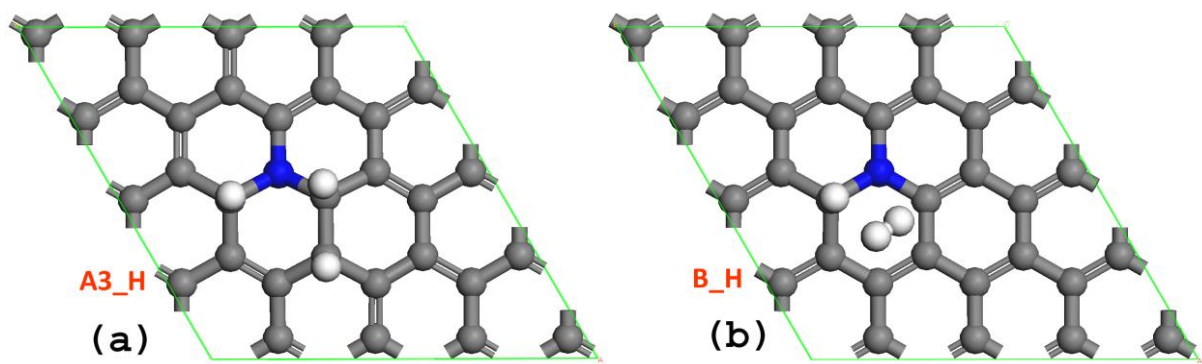

Fig. 6

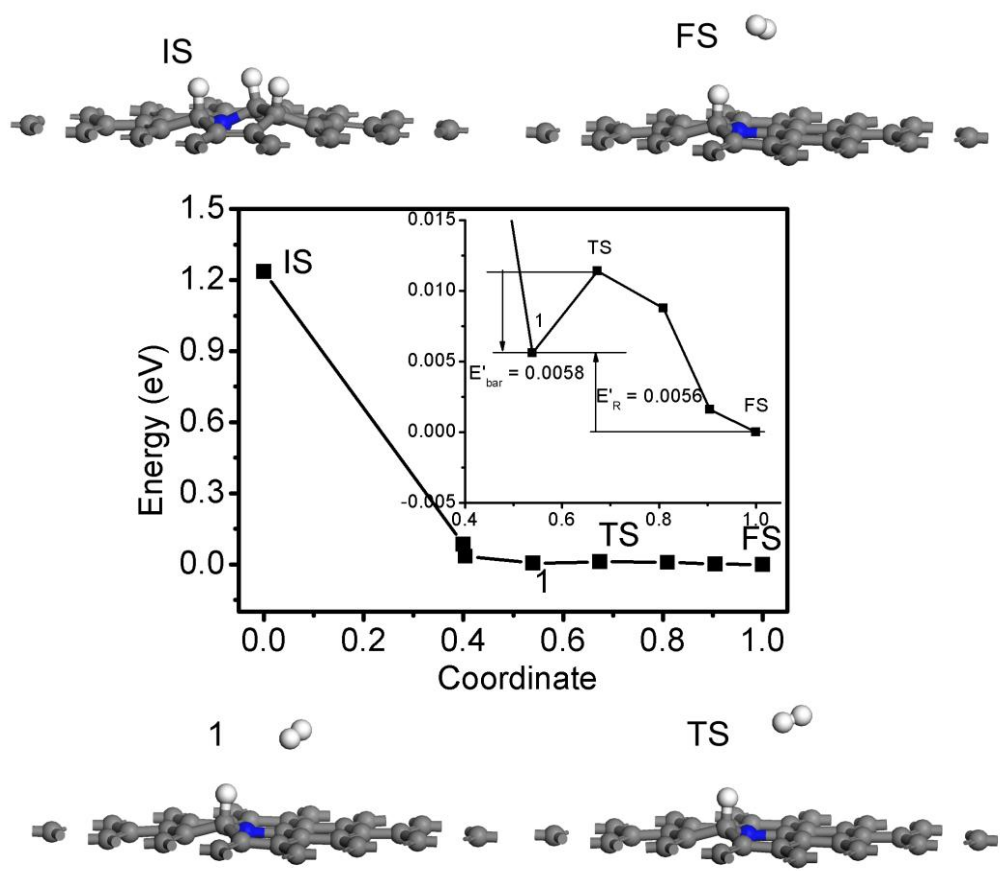